\documentclass[mathpazo]{cicp}

\def\pd#1#2{ \frac{\partial #1}{\partial #2}}
\def\bvec{\left\{\begin{array}{c}} 
\def\evec{\end{array}\right\}}

\usepackage{graphicx}
\usepackage[usenames,dvipsnames,svgnames,table]{xcolor}

\newcommand\solidrule[1][1cm]{\rule[0.5ex]{#1}{.4pt}}

\newcommand\dotdashedrule{\mbox{%
  \solidrule[4mm]\hspace{0.8mm}\solidrule[0.3mm]\hspace{0.8mm}\solidrule[4mm]}}
\newcommand\verythindashedrule{\mbox{%
  \solidrule[0.65mm]\hspace{0.65mm}\solidrule[0.65mm]\hspace{0.65mm}\solidrule[0.65mm]\hspace{0.65mm}\solidrule[0.65mm]\hspace{0.65mm}\solidrule[0.65mm]\hspace{0.65mm}\solidrule[0.65mm]\hspace{0.65mm}\solidrule[0.65mm]}}
\newcommand\thindashedrule{\mbox{%
  \solidrule[1mm]\hspace{1mm}\solidrule[1mm]\hspace{1mm}\solidrule[1mm]\hspace{1mm}\solidrule[1mm]\hspace{1mm}\solidrule[1mm]}}
\newcommand\fewdots{\mbox{%
   \hspace{0.25mm} \texttt{o} \hspace{0.25mm}}}

\begin{document}

\title{A Modified Gas-Kinetic Scheme for Turbulent Flow}
\author[M. Righi]{Marcello Righi\corrauth}
\address{School of Engineering, Zurich University of Applied Sciences, Technikumstrasse 9, 8401 Winterthur, Switzerland}
\email{{\tt marcello.righi@zhaw.ch} }

\begin{abstract}

By analogy with the kinetic theory of gases, most turbulence modeling strategies rely on an eddy viscosity to model the unresolved turbulent fluctuations.
However, the ratio of unresolved to resolved scales - very much like a degree of  rarefaction - is not taken into account by the popular conventional schemes, 
based on the Navier-Stokes equations.
This paper investigates the simulation of turbulent flow with a gas-kinetic scheme. 
In so doing, the modeling of turbulence becomes part of the numerical scheme: 
the degree of rarefaction is automatically taken into account; 
the turbulent stress tensor is not constrained into a linear relation with the strain rate.
Instead it is modeled on the basis of the analogy between particles and  eddies, without any assumptions on the type of turbulence or flow class. 
The implementation of a turbulent gas-kinetic scheme into a finite-volume solver is put forward, with 
turbulent kinetic energy and dissipation supplied by an allied turbulence model. 
A number of flow cases, challenging for conventional RANS methods, are investigated;   
results show that the shock-turbulent boundary layer is  captured in a much more convincing way by the gas-kinetic scheme.

\end{abstract}

\pac{47.27.em, 47.27.nb, 47.32.Ff, 47.40.Hg, 47.40.Nm 	, 47.45.Ab, 47.85.Gj, 47.85.ld}

\maketitle

\section{Introduction}
\label{sec:intro}

A number of gas-kinetic schemes have been developed over the latest 20 years (refer to \cite{xu2001gas,may2007improved,mandal1994kinetic,chou1997kinetic,xu1994numerical} and references therein).
With respect to conventional CFD, these schemes, derived from the Boltzmann equation,  exploit a physically more consistent model of fluid mechanics.
Gas-kinetic schemes are more accurate, 
in particular in the presence of shock waves. 
Besides, they are more suitable to high-order reconstruction (refer to \cite{li2010high,xu2005multidimensional,xuan2012new} and references therein) and 
may be used as a platform to investigate rarefied flow: either, in the transitional regime, by relating the relaxation time to macroscopic variables and their gradients (refer to Liao {\sl et al.} \cite{liao2007gas})  or introducing additional discrete velocity levels (refer to Xu {\sl et al.} \cite{xu2010unified}).  

The most evident characteristics of these schemes concerns the handling of collisions. 
Whereas in conventional schemes, collisions are modeled by a viscosity and assuming a linear stress - strain relation, gas kinetic schemes use a relaxation time - related to collision frequency and hence to viscosity. 
The BGK model, depending on the scheme parameters, generates a number of different contributions: a stress tensor proportional to viscosity and strain rate,
but also a number of non-linear correction terms, related to the local degree of rarefaction and the macroscopic gradients (refer to Xu \cite{xu2001gas} and to May {\sl et al.} \cite{may2007improved} for additional analysis). These correction terms represent the ``kinetic effects'' that affect the transport properties of the flow, as soon as the degree of rarefaction becomes significant.
In most of the flows of engineering and scientific interest, rarefaction may become significant only in shocklayers; however, even in the presence of shocks, conventional schemes do not correct advection: the solution of the Riemann problem is unaffected by the physical viscosity of the flow.

When it comes to turbulence modeling, conventional schemes express the turbulent stress tensor in a strain rate series, where the first terms is normally retained and the second only in the case of non-linear turbulence models (refer for instance to Pope \cite{pope2000turbulent} and to Chen {\sl et al.} \cite{chen2004expanded} for a discussion of the link between non-linear models and kinetic theory).
Some of these models are successful for a number of flow classes. However, the closures are often specific to a flow class.
Using a  gas-kinetic scheme for turbulent flow relies on the analogy between particles and turbulent eddies, which is anyway the basis for the concept of eddy viscosity 
(refer to Chen {\sl et al.} \cite{chen2003extended} 
for a more detailed discussion). The dynamics of unresolved turbulence  is modeled by a turbulent relaxation time - supplied by an allied turbulence model integrated in time alongside the macroscopic variables. The degree of rarefaction would be measured by the ratio of unresolved turbulent fluctuations to the resolved scales. 
The turbulent stress tensor is not simply linearly related to strain: the scheme  modifies it, through the non-linear correction terms, without any assumptions on the nature of turbulence. 

Gas-kinetic schemes, unlike Lattice Boltzmann methods, have not been used systematically to simulate turbulent flow. This paper investigates the adaptation of a well-validated 
gas-kinetic scheme, the one published  by Xu in 2001 \cite{xu2001gas}, to the simulation of turbulence with a simple RANS approach; it does not propose a novel turbulence model nor a new modeling technique. Instead, a turbulent relaxation time is derived from the turbulent quantities provided by a standard two-equation turbulence model. The analysis of the resulting turbulent scheme shows that the turbulent stress tensor is corrected as a function of the degree of rarefaction, which  may assume ``transitional'' values in shocklayers. 
The results of numerical experiments, based on different examples of shock-turbulent boundary layer interaction, confirm that the gas-kinetic scheme has the ability to do better than conventional schemes. 

This paper is structured as follows: the derivation of the standard, laminar, gas-kinetic scheme and the modification into the turbulent gas-kinetic scheme are presented in section \ref{sec:turbulentgks}, section \ref{sec:numerical} is dedicated to numerical experiments; conclusions are presented in  section \ref{sec:conclusions}.

\section{A Gas-Kinetic Scheme for Turbulent Flow}
\label{sec:turbulentgks}

\subsection{Derivation of the gas-kinetic scheme }
\label{subsec:derivation}
%

\subsubsection{BGK model, derivation of Euler and Navier-Stokes equations}
\label{subsubsec:bgkmodel}

The kernel of a gas-kinetic scheme consists in modeling the fluxes of the conservative variables across computational cells on the basis of the Boltzmann equation,  instead of of the Navier-Stokes equations.
In practice, the collision operator in the Boltzmann equation is linearized, often according to the BGK model \cite{bhatnagar1954model}: 

\begin{equation}
\pd{f}{t} + (u \cdot \nabla) f = \frac{f^{eq}-f}{\tau},
\label{eq:bgk}
\end{equation}

\noindent where the relaxation time $\tau$ is related to the frequency of collisions and $f^{eq}$ is a Maxwell-Boltzmann distribution function, describing a gas in thermodynamic equilibrium: 

\begin{equation}
f^{eq} = \rho \left( \frac{\lambda}{\pi} \right)^{\frac{N+2}{N}} \exp \left[{ -\lambda
\left(
(u_i-v_i)^2+\xi^2 \right)} \right].
\label{eq:maxwellian}
\end{equation}

In Eqq. \ref{eq:bgk} and \ref{eq:maxwellian}, where summation convention holds, and in the rest of this section, the macroscopic variables density, velocity and total energy are indicated with $\rho$, $v$ and $E$; the microscopic variables are velocity $u$ and the $N$ effective degrees of freedom of the gas molecules $\xi$. The quantity of effective degrees of freedom is  $N=\left({5-3\gamma}\right)/\left({\gamma-1}\right)+1$,
where $\gamma$ is the specific heat ratio. 
Moreover, $\lambda = {m}/\left({2kT}\right) = \rho/2 p$, $m$ is the molecular mass, $k$ is the Boltzmann constant, and $T$ is temperature.

The conservative variables $w$ can be recovered by taking moments of the distribution function: 

\begin{equation}
w = 
\int\psi f d\Xi,
\label{eq:moment1} 
\end{equation}

\noindent where the elementary volume in phase space is $d\Xi = du_1 du_2 du_3 \, d\xi$ and:

\begin{equation}
\psi = 
\left[
1 \;\;\,\,
v_1 \;\;\,\,
v_2 \;\;\,\,
v_3 \;\;\,\,
\frac 12 \left( {u_i}^2 +\xi^2 \right) 
\right]^T.
\end{equation}

In order to be practically used in the development of a numerical scheme Eq. \ref{eq:bgk} is expanded according to the  the Chapman-Enskog method (refer to Cercignani \cite{cercignani1988boltzmann} and Xu \cite{xu1998gas} for a formal discussion); by introducing the non-dimensional quantity $\epsilon = \tau / \widehat{\tau}$, where $\widehat{\tau}$ is a reference time scale, Eq. \ref{eq:bgk} is re-written in the form: 

\begin{equation}
f = f^{eq} - \epsilon \widehat{\tau} D f,
\label{eq:chapman-enskog0}
\end{equation}

\noindent with $D \cdot = \pd{}{t} \cdot + (u \cdot \nabla) \cdot $. By substituting Eq. \ref{eq:chapman-enskog0} into the right hand side of the same equation, one obtains: 

\begin{equation}
f = f^{eq} - \epsilon \widehat{\tau} D f^{eq} + \epsilon^2 \widehat{\tau} D \left( \widehat{\tau} D f^{eq} \right) + \dots. 
\label{eq:chapman-enskog}
\end{equation}
The Euler and Navier-Stokes equations can be obtained by taking moments of Eq. \ref{eq:chapman-enskog}:

\begin{equation}
\int f d\Xi = \int \left( f^{eq} - \epsilon \widehat{\tau} D f^{eq} + \epsilon^2 \widehat{\tau} D \left( \widehat{\tau} D f^{eq} \right) + \dots \right) d\Xi. 
\label{eq:moments-chapman-enskog}
\end{equation} 
Assuming that  $\epsilon$ is a small quantity, the expansion in Eq. \ref{eq:moments-chapman-enskog} can be truncated.
It can be demonstrated (the complete derivation can be found in Cercignani \cite{cercignani1988boltzmann} and Xu \cite{xu1998gas}) that 
Eq. \ref{eq:moments-chapman-enskog} corresponds to the Euler equations if the terms ${\cal O} (\epsilon)$ are dropped, and to the Navier-Stokes equations if the terms 
${\cal O} (\epsilon^2)$ are dropped.
The conditions to recover the Navier-Stokes equations for a diatomic gas are:
 
\begin{itemize}
\item $\mu = \tau/p$ (bulk viscosity is $2N/(3K+9) \mu$) 
\item $Pr= \mu C_p / \kappa = 1$, this being  a known drawback of the BGK model. 
\end{itemize}
Distribution functions at Euler and Navier-Stokes level are therefore:

\begin{eqnarray}
f^{Euler} &=& f^{eq}, 
\label{eq:chapman-enskog-euler} \\ 
f^{NS}  &=& f^{eq} - \epsilon \widehat{\tau} D f^{eq}. 
\label{eq:chapman-enskog-ns}
\end{eqnarray}

\subsubsection{Model of the flow at cells interface}
\label{subsubsec:flowinterface}

At each interface, at the beginning of a time step, a function $f$ is introduced as a solution to the Boltzmann equation, with initial conditions consistent with the gas states at both sides of the interface. 
A closed-form solution of the BGK equation in a time interval $[0,t]$ is given by the integral form presented by Kogan \cite{kogan1969rarefied} and is used to evaluate $f$:


\begin{eqnarray}
f^{BGK}(x_1,x_2,x_3,t,u_1,u_2,u_3,\xi) &=& 
\frac{1}{\tau} \int_o^t f^{eq}(x_1',x_2',x_3',t,u_1,u_2,u_3,\xi) e^{-(t-t')/\tau}\,dt' \nonumber \\ 
                                       &+& e^{-t/\tau} f_0 (x_1 - u_1 t,x_2 - u_2 t,x_3 - u_3 t), 
\label{eq:integralbgk}
\end{eqnarray}
 
\noindent where $x_1' = x_1 - u_1 (t-t'),\,\,\,x_2'=x_2-u_2 (t-t'),\,\,\,x_3'=x_3-u_3(t-t')$. 

%
%

For the sake of clarity, in the rest of this section, the interface is assumed to be normal to $x_1$ which is indicated with the symbol $x$ to reduce the number of indexes,
the microscopic velocity $u_1$ is indicated with $u$. 
The left and right states of the gas are indicated with the suffixes $(l)$ and $(r)$. A third, fictitious  gas state representing the gas at the interface is indicated with the suffix $(c)$. The left and right values of the conservative variables $w_{(l)}$, $w_{(r)}$ and their gradients ${w_{(l)}}_{/x}$, ${w_{(r)}}_{/x}$ are obtained from a standard reconstruction scheme and limiting process (e.g. MUSCL/TVD, ENO, WENO).
On both sides a Maxwell-Boltzmann distribution is defined, from  $w_{(l)}$ and $w_{(r)}$:

\begin{equation}
f^{eq}_{(l)} = \rho_{(l)} \left( \frac{\lambda_{(l)}}{\pi} \right)^{\frac{N+2}{N}} \exp \left[{ -\lambda_{(l)}
\left(
(u_i-{v_{(l)}}_i)^2+\xi^2 \right)} \right],
\label{eq:lmaxwellian}
\end{equation} 

\begin{equation}
f^{eq}_{(r)} = \rho_{(r)} \left( \frac{\lambda_{(r)}}{\pi} \right)^{\frac{N+2}{N}} \exp \left[{ -\lambda_{(r)}
\left(
(u_i-{v_{(r)}}_i)^2+\xi^2 \right)} \right].
\label{eq:rmaxwellian}
\end{equation} 
The intermediate state is reconstructed from the left and right states:

\begin{equation}
w_{(c)} = \int_{u<0} h^l f^{eq}_{(l)} \psi d\Xi + \int_{u>0} h^r f^{eq}_{(r)} \psi d\Xi,
\label{eq:intermediatestate}
\end{equation}
\noindent where  $h^l= H(u)$, $h^r = 1-H(u)$ and $H$ is the Heaviside function.
A third Maxwell-Boltzmann distribution is then defined on the basis of $w_{(c)}$:

\begin{equation}
f^{eq}_{(c)} = \rho_{(r)} \left( \frac{\lambda_{(r)}}{\pi} \right)^{\frac{N+2}{N}} \exp \left[{ -\lambda_{(c)}
\left(
(u_i-{v_{(c)}}_i)^2+\xi^2 \right)} \right].
\label{eq:cmaxwellian}
\end{equation}

The distribution functions $f_0$ and $f^{eq}$ in Eq. \ref{eq:integralbgk} can be defined on the basis of these three states. The initial condition $f^{BGK}(x,v,0) = f_0$ is defined as a solution to the BGK model \ref{eq:bgk} at Navier-Stokes level. $f_0$ is obtained from Eq. \ref{eq:chapman-enskog-ns}, imposing a discontinuity between left and right states and truncating the Chapman-Enskog expansion at the second term as in Eq. \ref{eq:chapman-enskog-ns}:    

\begin{equation}
f_0= \left\{\begin{array}{l}
{f^{eq}_{(l)}} \left( \left(1+{a_{(l)}} x \right)-\tau \left( {a_{(l)}} u+A_{(l)} \right) \right),\,\,\,x \leq 0, \\
{f^{eq}_{(r)}} \left( \left(1+{a_{(r)}} x \right)-\tau \left( {a_{(r)}} u+A_{(r)} \right) \right),\,\,\,x >0, \end{array} \right. 
\label{eq:initialf0}
\end{equation}

\noindent where $a_{(l)}$ and $a_{(r)}$ are the coefficients of spatial expansion in the phase space,
$A_{(l)}$ and $A_{(r)}$ are the first order coefficient of the temporal expansions. 
The coefficients $a_{(l)}$ and $a_{(r)}$  may be calculated from the gradients of the conservative variables. 
Although it is not shown here, $a_{(l)}$ and $a_{(r)}$ are not constant values but approximated as linear functions of all degrees of freedom of the gas (microscopic velocities $u_i$ and internal effective degrees of freedom $\xi$).
The coefficients $A_{(l)}$ and $A_{(r)}$  are not calculated from the past history of the flow, but from the compatibility condition  $\int ( f^{eq} - f ) \psi d\Xi = 0$ at $t=0$.



The equilibrium distribution approach by $f^{BGK}$ in Eq. \ref{eq:integralbgk} is expressed as: 

\begin{equation}
f^{eq} = \left\{
\begin{array}{l}
f^{eq}_{(c)} \left(1+\overline{ a}_{(l)} x-\overline{ A} t \right),\,\,\,x \leq 0, \\
f^{eq}_{(c)} \left(1+\overline{ a}_{(r)} x-\overline{ A} t \right),\,\,\,x >0, 
\end{array} 
\right. 
\label{eq:eqlgform}
\end{equation}

\noindent where the coefficient $\overline{ a}_{(l)}$ and  $\overline{ a}_{(r)}$ are obtained from fictitious gradients from the linear interpolation between $w_{(l)}$, $w_{(c)}$ and $w_{(r)}$. 
$\overline{A}$ are obtained from the compatibility condition integrated in a time interval. 

The substitution of Eqq. \ref{eq:initialf0} and \ref{eq:eqlgform} into Eq. \ref{eq:integralbgk} finally provides the solution to the BGK equation $f^{BGK}$:

\begin{eqnarray}
f^{BGK}  &=& \left[
       \left( 1 - e^{-t/\tau} \right) 
  + u \left( - \tau  + \tau  e^{-t/\tau}  +t\,e^{-t/\tau} \right) 
  \left( h_{(l)} \, \overline{a}_{(l)} +  h_{(r)} \, \overline{a}_{(r)} \right)    
  + \left( t - \tau  + \tau e^{-t/\tau}  \right) \overline{A} \right] f^{eq}_{(c)} 
  \nonumber \\
   &+& e^{-t/\tau}  \left[  h_{(l)} f^{eq}_{(l)} + h_{(r)} f^{eq}_{(r)} 
   - u \left( t + \tau\right)  \left(  a_{(l)} h_{(l)} f^{eq}_{(l)} + a_{(r)}^r h_{(r)} f^{eq}_{(r)}  \right) \right] \nonumber \\ 
   &-& \tau  e^{-t/\tau}  \left( A_{(l)} h_{(r)} f^{eq}_{(l)} + A_{(r)} h_{(r)} f^{eq}_{(r)} \right)    
.
\label{eq:resultingf}
\end{eqnarray}
In order to obtain a more compact formulation, the following distribution functions at Navier-Stokes level are introduced: 

\begin{eqnarray}
f_{(c)} &=& 
f^{eq}_{(c)} \left( 1 - \tau \left( h_{(l)} \overline{a}_{(l)} u + h_{(r)} \overline{a}_{(r)} u + \overline{A} \right) \right), 
\label{eq:fcentral} \\
f_{(u)} &=& 
  h_{(l)} f^{eq}_{(l)}  \left[ 1 - \tau \left(  a_{(l)} u + A_{(r)} \right) \right] 
+ h_{(r)} f^{eq}_{(r)}  \left[ 1 - \tau \left(  a_{(r)} u + A_{(r)} \right) \right]. 
\label{eq:fupwind}
\end{eqnarray}


\noindent $f_{(c)}$ is built from the fictitious state  introduced with Eq. \ref{eq:intermediatestate} and, by analogy with the terminology used for numerical schemes, may be considered {\sl central}. 
$f_{(u)}$ keeps into account  the left and right reconstructed variables, and may be related to the idea of {\sl upwind}. The use of the terms {\sl central} and {\sl upwind} 
do not imply any analogy with conventional schemes involving a discontinuous reconstruction, where {\sl upwind} in included to generate dissipation. 
When combined with Eqq. \ref{eq:fcentral} and \ref{eq:fupwind}, Eq. \ref{eq:resultingf} can be re-expressed:

\begin{equation}
f^{BGK} = f_{(c)} ( 1 + \bar{A} t ) +
e^{-t/\tau} \left( f_{(u)} - f_{(c)} \right) +
t e^{-t/\tau} \left( \widetilde{f_{(u)}} - \widetilde{f_{(c)}} \right),
\label{eq:resultingf2}
\end{equation}

\noindent where $\widetilde{f_{(c)}}$ and $\widetilde{f_{(u)}} $ only retain the spatial expansion coefficients: 

\begin{eqnarray}
\widetilde{f_{(c)}} &=& 
f^{eq}_{(i)} \left( 1 - \tau \left( h_{(l)} \overline{a}_{(l)} u + h_{(r)} \overline{a}_{(r)} u  \right) \right), 
\label{eq:fcentraltilde} \\
\widetilde{f_{(u)}} &=& 
  h_{(l)} f^{eq}_{(l)}  \left[ 1 - \tau \left(  a_{(l)} u  \right) \right] 
+ h_{(r)} f^{eq}_{(r)}  \left[ 1 - \tau \left(  a_{(r)} u  \right) \right]. 
\label{eq:fupwindtilde}
\end{eqnarray}
Eq. \ref{eq:resultingf2} reveals a combination of {\sl central} and {\sl upwind} distribution functions, whose coefficients depend on collision rate $\tau$ and time. 
The limit of Eq. \ref{eq:resultingf2}  for a vanishing $\epsilon$ (or {\sl hydrodynamic limit}) is: 

\begin{equation}
\lim_{\epsilon \mapsto 0} f^{BGK} = f_{(c)} ( 1 + \bar{A} t ).
\label{eq:hydrodynamiclaminar}
\end{equation} 
Eq. \ref{eq:hydrodynamiclaminar} suggests that the gas-kinetic scheme for small values of $\epsilon$ generates time-accurate Navier-Stokes fluxes by means of $f_{(c)}$.
Non negligible values of $\epsilon$ trigger corrections depending on reconstruction values and the degree of rarefaction $\epsilon$.
This implies the capability to generate  physically consistent dissipation as a reaction to a discontinuity in the reconstruction:

\begin{equation}
f^{BGK}_{corr} = 
e^{-t/\tau} \left( f_{(u)} - f_{(c)} \right) +
t e^{-t/\tau} \left( \widetilde{f_{(u)}} - \widetilde{f_{(c)}} \right).
\label{eq:correctionterms}
\end{equation}
Eq. \ref{eq:resultingf2} can then be re-expressed as: 

\begin{equation}
f^{BGK}=f^{NS}+f^{BGK}_{corr} (\epsilon).
\label{eq:fnspluscorrections}
\end{equation}
It is interesting to note that the correction terms generate viscous stresses which are in principle not linearly related  to the strain tensor. 
The stress tensor  can be evaluated by integrating the fluctuations (refer to Xu \cite{xu1998gas}):

\begin{equation}
\sigma^{BGK}_{ij} = \int (u_i -v_i)(u_j - v_j) \, f^{BGK} \, d\Xi.
\label{eq:stresstensorgks}
\end{equation}
One can now decompose the stress tensor into a Navier-Stokes component and a correction term:

\begin{equation}
\sigma^{BGK}_{ij} = \sigma^{NS}_{ij} + {\sigma^{BGK}_{corr}}_{ij} = \int (u_i -v_i)(u_j - v_j) \, f^{NS} \, d\Xi  + \int (u_i -v_i)(u_j - v_j) \, f^{BGK}_{corr} \, d\Xi.
\label{eq:stresstensorgks2}
\end{equation}
It can be easily demonstrated (Xu \cite{xu1998gas}) that $f^{NS}$ generates the ``conventional'' Navier-Stokes stress tensor:

\begin{equation}
{\sigma^{NS}_{ij}} =  -p \delta_{ij} + \mu \left( {S_{(c)}}_{ij} - \frac 23 {S_{(c)}}_{kk} \delta_{ij} \right),
\label{eq:stresstensor1}
\end{equation}

\noindent where:

\begin{equation}
{S_{(c)}}_{ij} = \frac12 \left( \pd{{v_{(c)}}_i}{x_j}+ \pd{{v_{(c)}}_j}{x_i} \right)
\label{eq:strainrate}
\end{equation}
is the strain rate tensor and $\mu = \tau / p$.  The suffix $(c)$ refers to the intermediate gas state.
Interestingly, ${\sigma^{BGK}_{corr}}_{ij}$ does not show any proportionality to a single strain tensor, as $f_{(u)}$ cannot be related to a single macroscopic state.

\subsubsection{Evaluation of fluxes in a time interval}
\label{subsubsec:fluxes}

The solution of the BGK equation Eq. \ref{eq:resultingf2} allows the recovery of the numerical fluxes in a given time interval $[0,T]$ by a simple time integration: 

\begin{equation}
{\cal{F}} = \int_0^{T} \int f^{BGK} \psi d\Xi \, dt.
\label{eq:timeint4fluxes}
\end{equation} 
In the original scheme by Xu \cite{xu2001gas}, the upper limit of the time interval, $T$, corresponds to the time step $T=\Delta t$.
This choice means that the solution of the BGK model is exploited only up to $\Delta t$, which is quite acceptable for laminar flow, for which the relaxation time $\tau$ is normally (depending on grid and flow) much smaller. 

Two arguments have suggested a modification to the original scheme: 
(i) when applying the gas-kinetic scheme to turbulent flow,  the laminar relaxation time $\tau$ must be replaced by a turbulent one, the ratio $\tau/\Delta t$ may assume much higher values and lead to numerical instability, and 
(ii) a dependence on the grid may be introduced in the solution  depending on time stepping technique and preconditioning. 

It is therefore proposed to introduce a new time scale $\widehat{\tau}$, representative of the resolved flow and grid independent. 
A common practice in rarefied flow dynamics is to evaluate $\widehat{\tau}$  on the basis of the resolved flow gradients, typically of density:

\begin{equation}
\widehat{\tau} = \frac{\rho}{D\rho}.
\label{eq:timescaleresolved}
\end{equation}
Time integration of Eq. \ref{eq:timeint4fluxes}  is now performed in the interval $[0,\widehat{\tau}]$:

\begin{equation}
{\cal{F}} = \int_0^{\widehat{\tau}} \int f^{BGK} \psi d\Xi \, dt = \widehat{\tau} \int \widehat{f^{BGK}} \psi d\Xi,
\label{eq:timeint4fluxes2}
\end{equation} 

\noindent where $\widehat{f^{BGK}}$ is the time average  of $f^{BGK}$:

\begin{eqnarray}
\widehat{f^{BGK}}&=& 
\frac{1}{\widehat{\tau}} \int _0^{\widehat{\tau}} f^{BGK}  dt = \nonumber \\ 
&=& f_{(c)} ( 1 + 1/2 \bar{A}\widehat{\tau} ) +
\frac{\tau}{\widehat{\tau}} \left( 1 - e^{-\widehat{\tau}/\tau} \right)                                                          \left( f_{(u)} - f_{(c)} \right) \nonumber \\
&+& 
\left[ \frac{\tau}{\widehat{\tau}}   \left( 1 - e^{-\widehat{\tau}/\tau} \right) - \frac{\tau}{\widehat{\tau}}  e^{-\widehat{\tau}/\tau} \right]   
\left( \widetilde{f_{(u)}} - \widetilde{f_{(c)}} \right).
\label{eq:averagef}
\end{eqnarray}
Introducing now the dimensionless quantities:

\begin{eqnarray}
\epsilon &=& \tau/\widehat{\tau}, 
\label{eq:epsilon1} \\ 
\alpha(\epsilon) &=& \epsilon \left( 1 - e^{-1/\epsilon} \right), 
\label{eq:alpha} \\ 
\beta(\epsilon) &=& \epsilon  e^{-1/\epsilon}, 
\label{eq:beta}
\end{eqnarray}

\noindent Eq. \ref{eq:averagef} can finally be re-arranged into the compact form: 

\begin{equation}
\widehat{f^{BGK}} = 
f_{(c)} ( 1 + 1/2 \bar{A} \widehat{\tau} ) +
\alpha  \left( f_{(u)} - f_{(c)} \right) +
\left(\alpha - \beta \right)  \left( \widetilde{f_{(u)}} - \widetilde{f_{(c)}} \right).
\label{eq:resultingf4}
\end{equation}

The dimensionless quantity $\epsilon$ is the ratio of unresolved time scales (thermal fluctuations in laminar flow) to the  ones of the resolved flow and represent a particular measure of the degree of rarefaction. 
Eq. \ref{eq:resultingf4} reveals therefore a dependence of the fluxes on the degree of rarefaction; contribution of ``upwind'' corrections are triggered by rarefaction.

In order to obtain numerical fluxes consistent with \ref{eq:resultingf4} over the time interval $[0,\Delta t]$, as is necessary in a practical calculation, an effective distribution function ${\overline{f^{BGK}}}$ is introduced:

\begin{equation}
{\cal{F}} =  \Delta t \int {\overline{f^{BGK}}} \psi d\Xi.
\label{eq:timeint4fluxes3}
\end{equation} 
${\overline{f^{BGK}}}$ is obtained by expanding  Eq. \ref{eq:timeint4fluxes2} in series and assuming a $\tau$ independent $\epsilon$:  

\begin{equation}
{\overline{f^{BGK}}} = 
f_{(c)} ( 1 + 1/2 \bar{A} \Delta t ) +
\alpha  \left( f_{(u)} - f_{(c)} \right) +
\left(\alpha - \beta \right)  \left( \widetilde{f_{(u)}} - \widetilde{f_{(c)}} \right).
\label{eq:resultingf5}
\end{equation}
The distribution function ${\overline{f^{BGK}}}$ in Eq. \ref{eq:resultingf5} is accurate to $\Delta t \tau$; the fluxes in Eq. \ref{eq:timeint4fluxes3} are accurate to  $\Delta t ^2 \tau$.

\subsection{Modification of the original scheme to simulate turbulent flow}
\label{subsec:turbulentscheme}

\subsubsection{Evaluation of the turbulent relaxation time}
\label{subsubsec:turbulenttime}

A gas-kinetic scheme for turbulent flow is obtained by replacing the relaxation time $\tau$  with a turbulent relaxation time $\tau_t$, 
representative of the dynamics of unresolved turbulence.
A turbulent relaxation time could be trivially derived from an assumed eddy viscosity $\mu_t$ by setting merely 
\begin{equation}
\tau_t = \mu_t /p,
\label{eq:tauttrivial}
\end{equation}

\noindent  by analogy with the relation $\tau=\mu/p$ used for laminar flow.

However, a deeper analysis of the scheme considers that the effect of unresolved turbulence is now expressed by a turbulent relaxation time and not by an eddy viscosity; $\tau_t$ can also be obtained in a more sophisticated and physically more meaningful way directly from assumed turbulent quantities like the turbulent kinetic energy $k$ and the turbulent dissipation rate $\varepsilon$. On the basis of a $k$-$\varepsilon$ RANS turbulence model and a systematic renormalization group procedure, Chen {\sl et al.}  \cite{chen2003extended} and Succi {\sl et al.} \cite{succi2002towards} proposed: 

\begin{equation}
\tau^{k\varepsilon}_t = \tau + C_{\mu}\frac{k^2/\varepsilon}{T\left( 1 + \eta^2 \right)^{1/2} } ,
\label{eq:tautchen}
\end{equation}

\noindent where $C_{\mu}$ is a numerical coefficient used in the $k$-$\varepsilon$ model, normally around $0.09$, $k$ is turbulence kinetic energy, $\varepsilon$ is turbulence dissipation rate and $\eta=S k/\varepsilon$, $S$ is a measure of the local velocity gradient. 
The argument used by Chen is that $\tau^{k\varepsilon}_t$ in  Eq. \ref{eq:tautchen} should express the dependence of $\tau_t$ from the variety of unresolved time scales.  

Eq. \ref{eq:tautchen} can be adapted to other turbulence models such as 
the well-known $k$-$\omega$ turbulence model by Wilcox \cite{wilcox2006}.  
The  turbulent relaxation time and the degree of rarefaction are calculated from Eq. \ref{eq:tautchen} keeping into account the definition of specific dissipation rate $\omega$ (Wilcox \cite{wilcox2006}): 

\begin{equation}
\tau^{k\omega}_t = \tau + \tau_a +\frac{k/\omega}{T\left( 1 + \eta^2 \right)^{1/2} }.
\label{eq:tautchenadjusted}
\end{equation}
The dimensionless quantity $\epsilon = \tau / \widehat{\tau}$ introduced in Eq. \ref{eq:epsilon1} becomes now a ``turbulent Knudsen number'', based on the ratio of unresolved, turbulent, to resolved time scales.



\subsubsection{Solution of the BGK model with a variable $\tau$}
\label{subsubsec:rarefied}

The use of Eq. \ref{eq:integralbgk} represents an approximation, as the relaxation time $\tau$ is in reality dependent on the flow gradients. This is not peculiar to turbulent flow if the dependence of molecular viscosity on temperature is taken into account with Sutherland's law.
It might be argued that the dynamics of the relaxation time is slower than the resolved flow. The validity of these assumption remains though to be consolidated. 

The turbulent gas-kinetic scheme, like the original laminar scheme, exploits the information contained in the measure of rarefaction $\epsilon$. 
It generates kinetic effects, by taking into account the effect of {\sl collisions} (in this case, between eddies) on {\sl transport}. 
In turbulent flow, $\epsilon$ assumes larger values, up to a few thousandths or a few hundredths in shocklayers at high Mach number, which would correspond to a flow in {\sl transitional } regime. 
Despite the fact that gas-kinetic schemes are not developed for rarefied flow, they might be able to handle moderate rarefaction, provided the collisions are suitably modeled. 
A similar application has been done in laminar flow;  Liao {\sl et al.}  \cite{liao2007gas}  have used the same gas-kinetic scheme with a variable relaxation time in order to model the collisions in the transitional regime. 

\subsubsection{Turbulent stress tensor}
\label{subsubsec:stresstensor}

In classical turbulence modeling the linear relation between turbulent stress and strain rate tensors 
is adequate for most types of flow but can become a ``constraint'' in particular cases. 
More sophisticated, and in principle also more accurate, models include non-linear components of the turbulent stress tensor. 
These components are supplied ad-hoc, with coefficients fixed in an {\sl a priori} fashion  or somehow related to the flow gradients (as in algebraic models for instance). These relations are derived from assumptions on the nature of turbulence and are not general.  
The turbulent gas-kinetic scheme does not make any assumptions on the type of turbulence nor does it introduce a fixed structure for the turbulent stress tensor. Still it has the ability to generate a non-linear relation between $\sigma_{ij}$ and $S_{ij}$. The non-linear, corrections terms expressed by Eq. \ref{eq:stresstensorgks2} meet the only requirement of being a solution of the BGK equation (Eq. \ref{eq:bgk}). 
This might be interpreted by saying that the mechanism behind the turbulent gas-kinetic scheme consists of supplying some ``degrees of freedom'' (in the case of Xu's scheme the left and right flow states and their gradients) and letting the underlying kinetic theory find the physically more consistent solution.

\subsection{Implementation of the turbulent gas-kinetic scheme into a finite-volume solver}
\label{subsec:implementation}

\subsubsection{Artificial dissipation }
\label{subsubsec:artificial}

In the presence of shocks, also in turbulent flow, it might necessary to include artificial dissipation, through an additional relaxation time component. The relaxation time is modified with the addition of a contribution ($\tau_a$) which provides additional dissipation in the proximity of discontinuities: 

\begin{equation}
\tau = \tau + \tau_t + \tau_a. 
\label{eq:relaxation time}
\end{equation}

\noindent The artificial dissipation time $\tau_a$, following Xu \cite{xu2001gas}, is modeled in a manner close to conventional CFD, i.e.  proportional to the pressure jump across the interface: 

\begin{equation}
\tau_a = C_a \frac{\left|p^r-p^l\right|}{\left|p^r+p^l\right|}\, \Delta t,
\label{eq:taua}
\end{equation}

\noindent where $p^l$ and $p^r$ are pressure values of the left and right states of the gas, $C_a$ is a coefficient.
$C_a$ plays a role in laminar simulations by providing numerical stability in the presence of strong shocks. Its value is often set in the $[0,1]$ range, depending on grid and reconstruction technique. In turbulent simulations its role is often less critical for numerical stability of the simulation  although it contributes to stabilizing the shock.  

\subsubsection{Unity Prandtl number}
\label{subsubsec:unity}

Complying with the original formulation of the scheme by Xu \cite{xu2001gas}, the heath flux must be corrected to account for a realistic Prantdl number.

\subsubsection{Multidimensional implementation }
\label{subsubsec:multidimensional}


In this section, the distribution function is expanded only in the direction $x$, normal to the interface, i.e. with the coefficients 
$a_{(l)}$, $a_{(r)}$, $\overline{a}_{(l)}$ and $\overline{a}_{(r)}$. This choice is computationally efficient and has also been made 
by other researchers  (refer for instance to Xu \cite{xu2001gas} and May  {\sl et al.} \cite{may2007improved}). 
However, the scheme can be easily modified in order to include a multi-dimensional expansion. Additional coefficients are needed for the other coordinate directions;  for instance Eq. \ref{eq:initialf0} would be re-expressed:

\begin{equation}
f_0= \left\{\begin{array}{l}
{f^{eq}_{(l)}} \left( \left(1+{a_{(l)}} x + {b_{(l)}} y \right)-\tau \left( {a_{(l)}} u_1+{b_{(l)}} u_2 + A_{(l)} \right) \right),\,\,\,x \leq 0, \\
{f^{eq}_{(r)}} \left( \left(1+{a_{(r)}} x + {b_{(r)}} y \right)-\tau \left( {a_{(r)}} u_1+{b_{(r)}} u_2 + A_{(r)} \right) \right),\,\,\,x >0. \end{array} \right. 
\label{eq:initialf0multi}
\end{equation}
A multi-dimensional version was also developed; on reasonably good-quality grids it provides comparable results at a slightly higher computational cost. 
On lower quality grids, the multi-dimensional version is prone to develop numerical instabilities.

\subsubsection{Allied turbulence model}
\label{subsubsec:allied}

Due to its higher suitability to the chosen flow cases, the well-known $k$-$\omega$ turbulence model has been chosen to supply the turbulent kinetic energy, $k$, and turbulent specific dissipation $\omega$.
The equations for  $k$ and $\omega$ are solved alongside the equations for the conservative variables. 
The complete formulation of the $k$-$\omega$ turbulence model can be found in Wilcox \cite{wilcox2006}. The version used is the one published in 2006.

\subsubsection{The complete turbulent gas-kinetic scheme}
\label{subsubsec:completetgks}

A turbulent gas-kinetic scheme based on the $k$-$\omega$ model is  represented by the following set of equations: 

\begin{eqnarray}
\epsilon^{k\omega} &=& \left( { \tau + \tau_a +\frac{k/\omega}{T \left( 1 + \eta^2 \right)^{1/2} } }\right) \left({\rho/D\rho}\right)^{-1},
\label{eq:epsilonkomega} \\
\alpha^{k\omega} &=& \epsilon^{k\omega}  \left( 1 - e^{-1/\epsilon^{k\omega} } \right), \; \; \beta^{k\omega} = \epsilon^{k\omega} \,  e^{-1/\epsilon^{k\omega} },   
\label{eq:alphabetakomega} \\
\overline{f^{k\omega}} &=& f^{(c)} ( 1 + 1/2 \bar{A} \Delta t ) +
\alpha^{k\omega} \left( f^{(u)} - f^{(c)} \right) +
\left(  \alpha^{k\omega} - \beta^{k\omega} \right)   \left( \widetilde{f^{(u)}} - \widetilde{f^{(c)}} \right), 
\label{eq:fkepsilon} \\
{\cal{F}} &=&  \Delta t \int {\overline{f^{k\omega}}} \psi d\Xi,
\label{eq:timeint4fluxes4}
\end{eqnarray}


\noindent where $\cal{F}$ are the numerical fluxes of the conservative variables. 

\subsubsection{Time advancing }
\label{subsubsec:time}

In the numerical experiments carried out in this work, the gas-kinetic fluxes have been implemented in a 2D finite volume steady-state solver (Righi \cite{righi2012aeronautical,righi2012rgd}). 
Well-known acceleration techniques (4-level multigrid and LU-SGS preconditioning in the form proposed by Jameson {\sl et al.} \cite{yoon1988lower,jameson1983solution}) have provided convergence properties comparable to more traditional Navier-Stokes schemes. As the LU-SGS preconditioning is hardly compatible with a parallelisation of the solver and cannot be considered a definitive solution.

\subsubsection{Variable reconstruction and limiting process}
\label{subsubsec:limiting}

The  reconstruction techniques include second and third order TVD/MUSCL  schemes and fifth order WENO, although the results shown are only second-order. The {\sl minmod} limiter has been used in all cases for both  conservative variables and their gradients.

\subsubsection{Boundary conditions}
\label{subsubsec:boundaryconditions}

No-slip wall boundary conditions have been used in all flow case; as no hypersonic cases have been simulated so far no changes with respect to conventional CFD have been implemented, as the ``rarefaction'' effects introduced in the previous sections are not expected to extend to solid walls.

\section{Numerical experiments }
\label{sec:numerical}

\subsection{Criteria for the selection of flow cases}
\label{subsec:criteria}

The interaction between a shock wave and a boundary layer originated by a compression corner, an impinging shock or an obstacle in a duct, develops into a multi-region flow each characterized by a different complex phenomenon (refer to \cite{babinsky2011shock,delery1986shock, dupont2006space, bookey2005experimental} for a detailed discussion of the flow physics involved). 
In particular, the adverse pressure gradients first causes the thickening of the boundary layer leading to a separation and to a large recirculation region. Turbulent stresses increase significantly downstream the shock wave originating a highly turbulent mixing layer in correspondence of the slip line. 

The flow cases have been chosen in order to evaluate the ability of the gas-kinetic scheme to cope with these different flow aspects and predict position and size of the interaction. 
Moreover the relatively rich set of experimental results available, including empirical scaling laws, allows evaluating the sensitivity to Reynolds number 
and geometric parameters. 
The analysis of the results refers to the well known AGARDograph by Delery {\sl et al.} \cite{delery1986shock}. The review paper  by Edwards \cite{edwards2008numerical} was also very useful. 
   
Beside the well-known D\'elery bump channel flow, three different cases of compression corner, at different angles, Reynolds and Mach number, and two cases of impinging shock have been chosen. 
Two of the compression corner cases refer to two experiments conducted at Princeton: the well-known 
measurements by Settles \cite{settles1976details}, conducted at a comparatively high Reynolds number, and the more recent ones by Bookey \cite{bookey2005experimental}, conducted at a much lower Reynolds number, to allow the comparison with DNS or LES numerical experiments.
The third compression corner concerns the experiments carried out by Dolling {\sl et al.} \cite{dolling1991unsteady} on a $28^\circ$ ramp at Mach $5$. 
The impinging shock cases refer to the experiments by Bookey {\sl et al.} \cite{bookey2005experimental} and by Dupont {\sl et al.} \cite{dupont2006space}.
The interaction is in this case very similar to the one caused by a compression corner with an angle $2\alpha$, when impinging shock is generated by a ramp of angle $\alpha$ (refer to D\'elery {et al.} \cite{delery1986shock}). 

It is well-accepted that the predictions obtained with RANS methods for these flow cases are generally inaccurate for one or more of the aspects mentioned above. Whereas algebraic stresses turbulence models provide better results than standard, linear two-equation models, the upstream influence distance (i.e. the distance between the corner and the point where the effect of the shock is felt by the flow) as well as the extension of the separated flow are often underpredicted; results obtained with conventional schemes can be found in \cite{edwards2008numerical,goldberg1998application,wallin2000explicit} and references therein, an excellent review can be found in the book by Babinsky and Harvey \cite{babinsky2011shock}.  
In practice, predictions in good agreements with experiments are obtained only with unsteady approaches such as DNS and LES \cite{pirozzoli2010direct, edwards2008numerical, garnier2009large}.


\begin{table}
\begin{tabular}{p{102mm}lrr}
\hline \hline
Case       & Reynolds$^1$ & Mach$^2$ \\ \hline
1.  D\'elery bump channel (Case C) \cite{delery1983experimental}   & $Re_{h}=1\;000\;000$   & $0.615$ \\   
2. Supersonic compression corner $\alpha = 8^\circ$, $16^\circ$, $20^\circ$, $24^\circ$ \cite{settles1979detailed} &  $Re_\theta = 23\;000 $ & $2.85$ \\   
3. Supersonic compression corner $\alpha = 24^\circ$ \cite{bookey2005experimental} &  $Re_\theta = 2\;400 $  & $2.90 $ \\   
4. Impinging shock $\alpha = 12^\circ$ \cite{bookey2005experimental} &  $Re_\theta = 2\;400$ & $2.90 $ \\   
5. Impinging shock $\alpha = 7^\circ$, $8^\circ$, $8.8^\circ$, $9.5^\circ$ \cite{dupont2006space} &  $Re_\theta = 6\;900 $ & $2.3$ \\   
6. Supersonic compression corner $\alpha = 28^\circ$ \cite{dolling1991unsteady} &  $Re_\delta = 877\;000 $ & $4.95$ \\   
\hline \hline
\end{tabular}
\begin{tabular}{p{140mm}}
{\footnotesize $^1$ $h$=bump height, $\theta$=momentum thickness, $\delta$ = displacement thickness; flow cases 2 and 4 also include a sensitivity analysis to Reynolds number.}  \\
{\footnotesize $^2$ D\'elery bump channel: freestream Mach = $0.615$, the flow reaches $M \simeq 1.45$ before the shock.} 
\end{tabular}
\caption{Summary of flow cases }
\label{tab:flowcases}
\end{table}

\subsection{Grid independence of results}

Grid-independence is one of the objectives pursued by the modification of the gas-kinetic scheme. 
Results obtained with a coarse, a medium and a fine grid are shown at least for each type of flow (bump channel, compression corner, impinging shock). Flow cases 2 and 3 have been also simulated on two grids, different in size and generated with different algorithms. 
All computational meshes are stretched to improve shock and boundary layer resolution, the latter is guaranteed by the placement of the first layer of cells within the laminar sublayer. 

%


\subsection{Flow cases}
\label{subsec:flowcases}

\subsubsection{Flow Case 1. D\'elery bump channel  }
\label{subsubsec:delery}

The experiment  transonic bump flow (Case C), experimentally investigated by D\'elery \cite{delery1983experimental} was designed to produce a strong shock - boundary layer interaction leading to a large flow separation. 
The  Mach $0.615$  duct flow impacts a ramp-semicircular bump mounted on one side of the channel, reaching approximately Mach $1.45$ before the shock. The shock - boundary layer interaction generates the typical $\lambda$ structure, with the separation starting  at the foot  of the first leg.     
It is well known that predicting the position of the separation point and the size of the separated area are challenging for standard, linear two-equation models as they  tend to delay separation and underpredict the extension of the separation.

In order to match the experimental position of the shock, the outlet pressure is heuristically adjusted. The two solvers  therefore use slightly different values of outlet pressure. The calculation is fully turbulent. 

\begin{figure}[h]
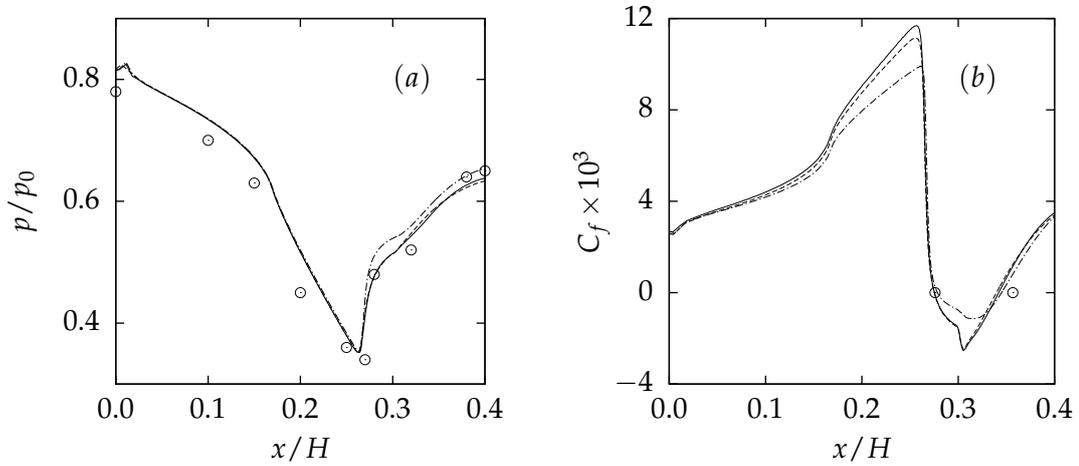

  \begin{center}
    \input{pdelery_}
    \input{cfdelery_}
  \end{center}
  \caption{ Pressure (a) and skin friction coefficient (b) for the D\'elery bump channel flow. 
   (\solidrule)  Gas-kinetic scheme (GKS) on finest grid,
   (\protect\thindashedrule) GKS on medium grid, 
   (\protect\verythindashedrule) GKS on coarsest grid,  
   (\protect\dotdashedrule) Navier-Stokes (Roe's approximate Riemann solver) on finest grid,
   (\protect\fewdots): experimental data from D\'elery \cite{delery1983experimental}.  
   Size of coarse, medium and fine grids:  $200 \times 160$, $312 \times 256$,  $456 \times 288$ respectively. 
   }
\label{fig:delerygridconvpcf}
\end{figure}

\begin{figure}[h!]
\begin{center}
\includegraphics[width=128mm]{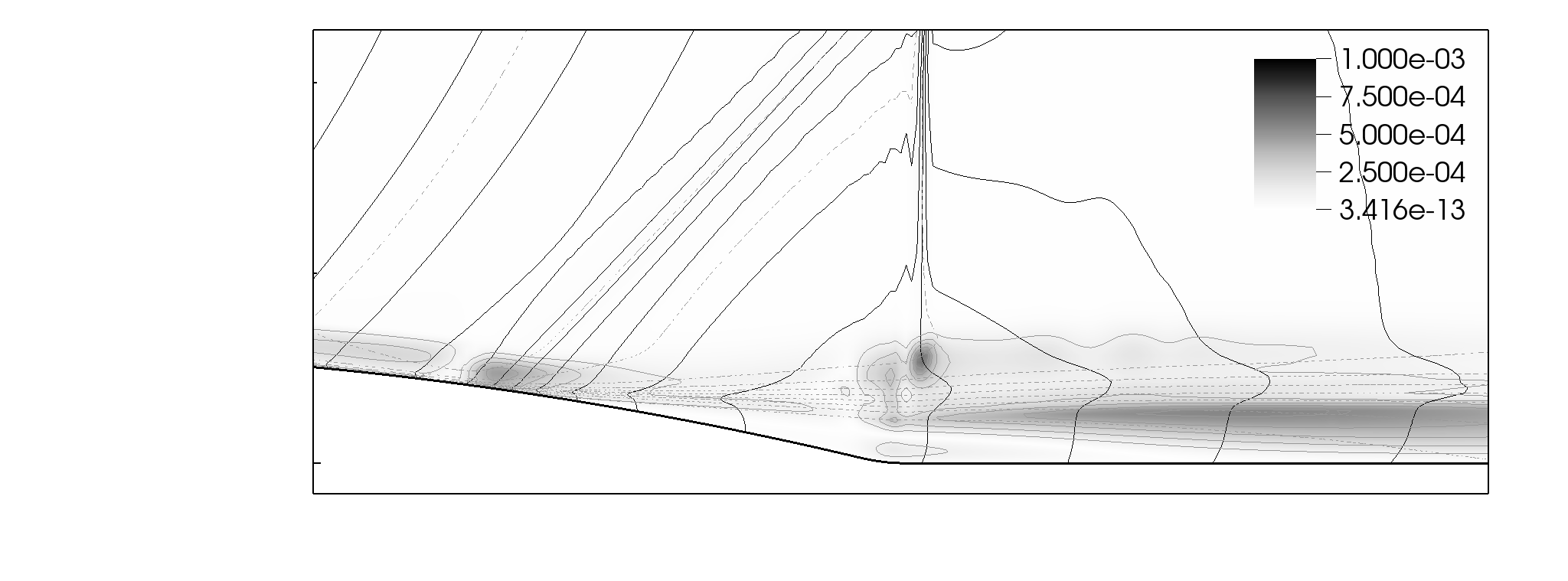}
\end{center}
\caption{Degree of rarefaction according to Eq. \ref{eq:epsilonkomega} (iso-contours in grayscale) for  D\'elery bump channel flow. 20 pressure isocurves have been added for reference. }
\label{fig:deleryrarefactionlevel}
\end{figure}

\begin{figure}[h!]
\begin{center}
\hspace{-18mm}
\includegraphics[width=84mm]{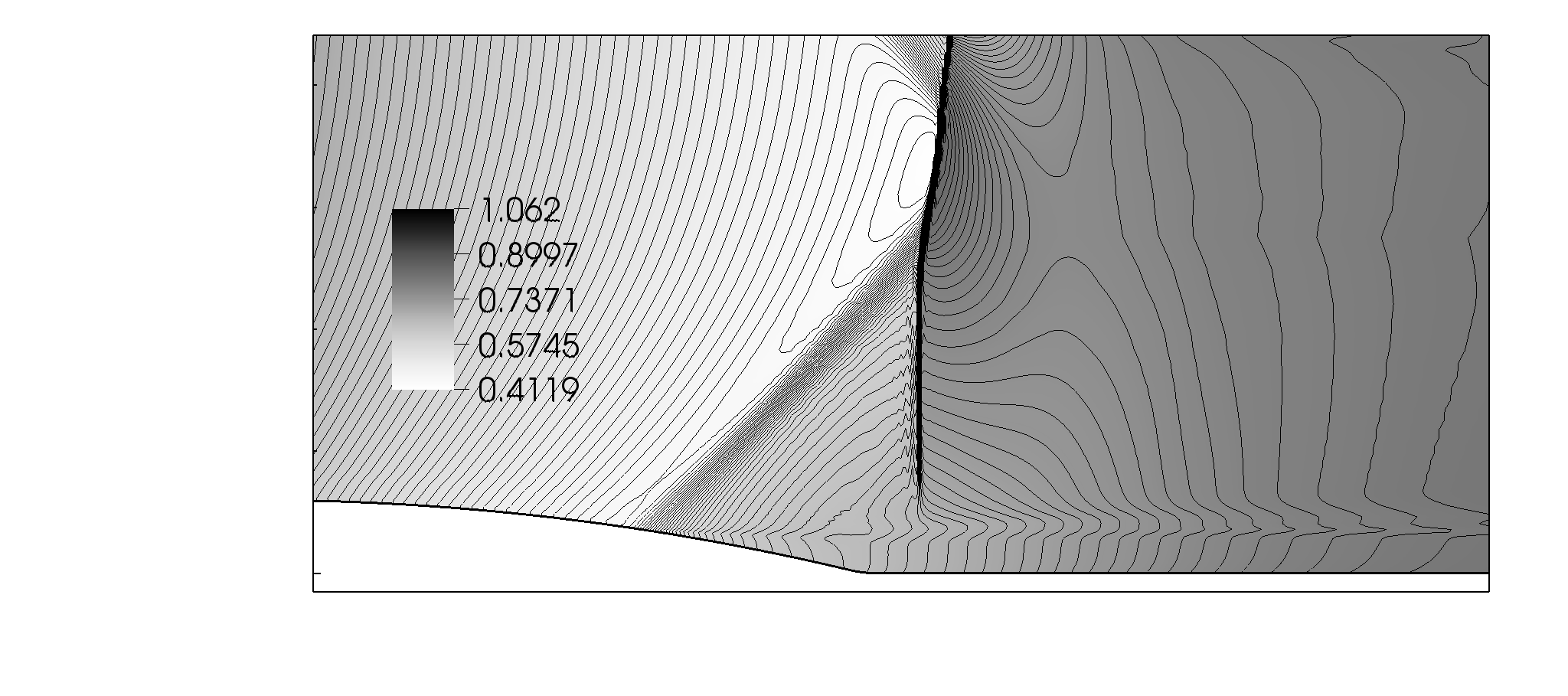}
\hspace{-5mm}
\includegraphics[width=84mm]{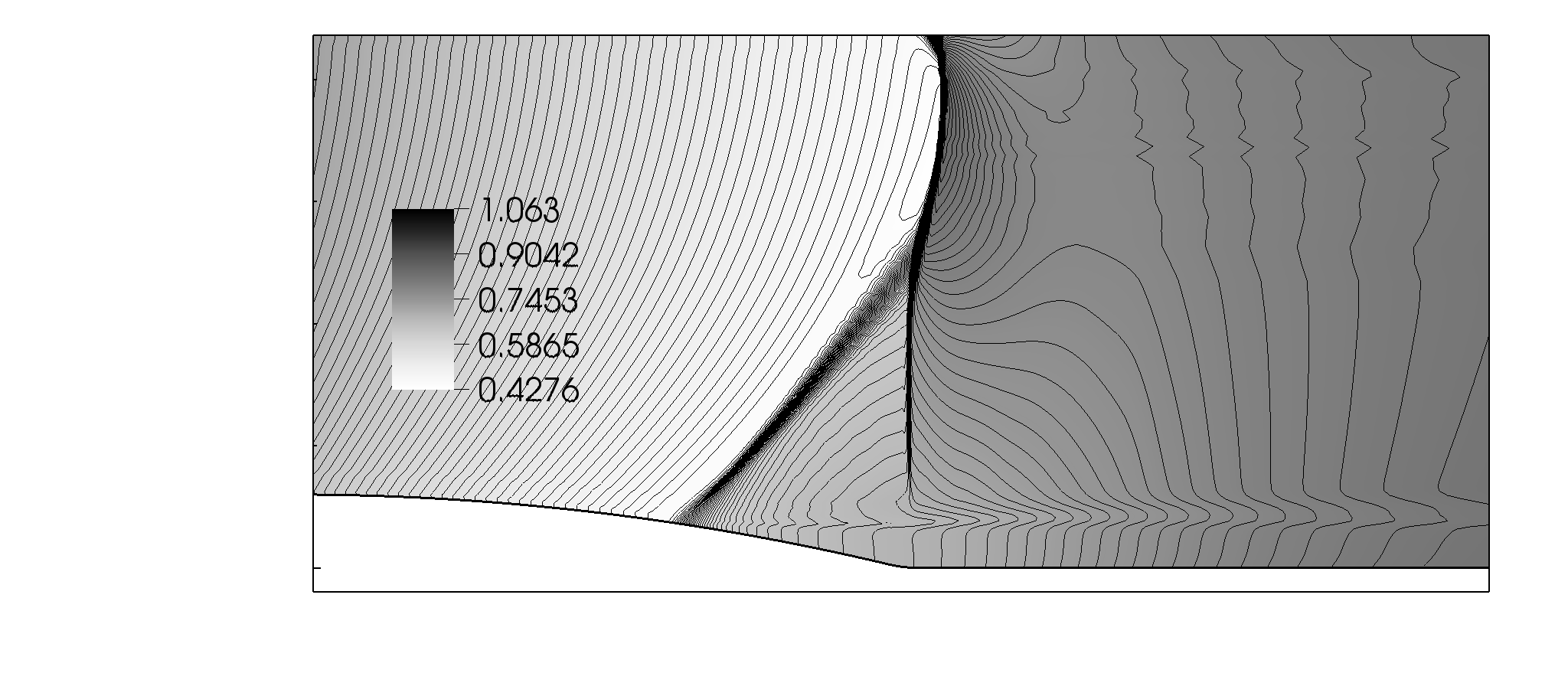}
\end{center}
\caption{Flow Case 1. Sketches of the shock system in  D\'elery bump channel flow. Gas-kinetic scheme on the left, Navier-Stokes on the right. In both figures, pressure is represented by iso-contours in grayscale. 100 pressure contour lines have also been added.  }
\label{fig:deleryshock}
\end{figure}

\begin{figure}[h!]
\begin{center}
\includegraphics[width=128mm]{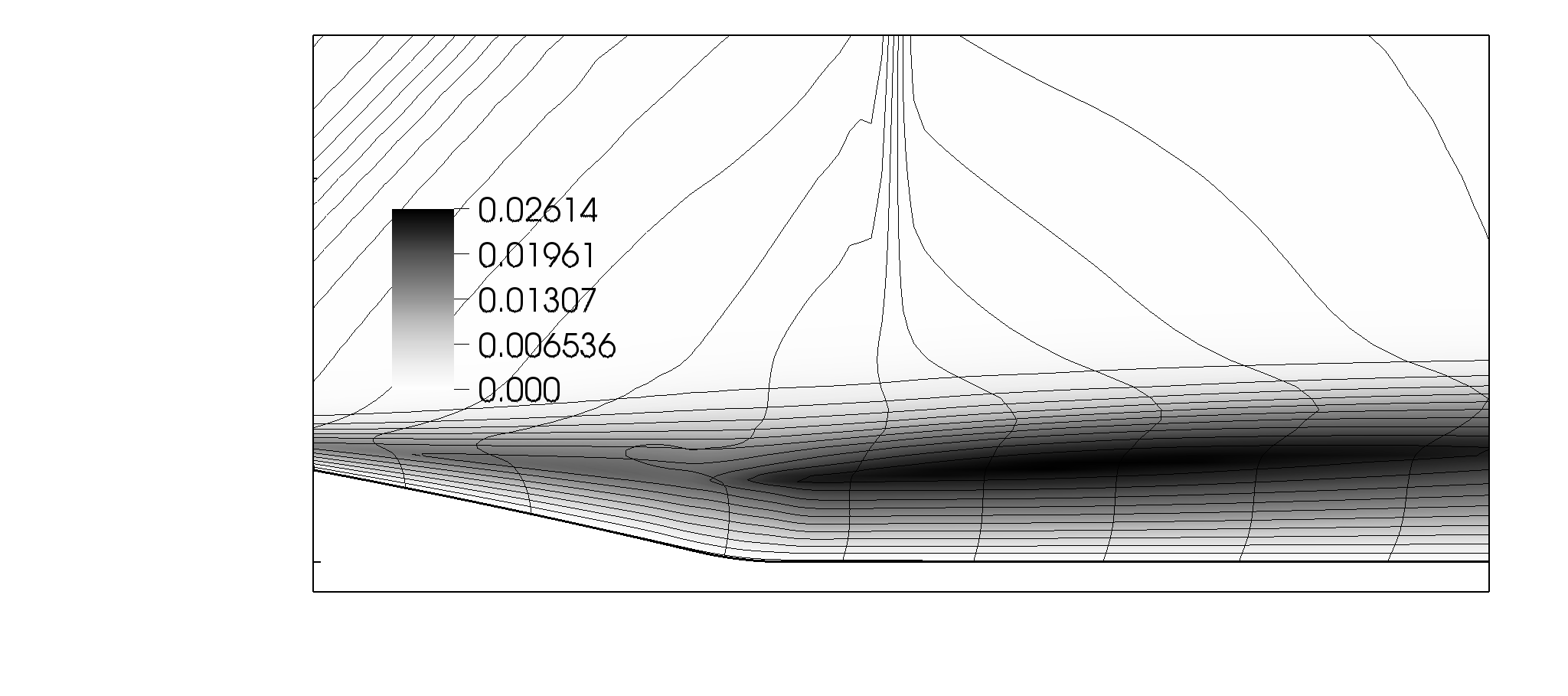}
\end{center}
\caption{Flow Case 1. Non-dimensional $\tau_{xy}$  component of the turbulent stress tensor in  D\'elery bump channel flow. Experimental values can for instance be found in the measurements by Sartor {\sl et al.} \cite{sartor2012piv}.  }
\label{fig:delerystressuv}
\end{figure}

Figures \ref{fig:delerygridconvpcf} to \ref{fig:delerystressuv} show the results of a simulation conducted with the turbulent gas-kinetic scheme and results obtained from the same solver but with a conventional  Navier-Stokes scheme with Roe's approximate Riemann solver.
Fig. \ref{fig:delerygridconvpcf} shows static pressure and skin friction coefficient. The predicted extension of the separation region is in good agreement with experimental data for both schemes, but the behavior of static pressure downstream of the shock is more accurately predicted by the gas-kinetic scheme. The predicted shock structure, shown in in Fig. \ref{fig:deleryshock}, is  different between the two schemes. One way to interpret this result, is that where the conventional scheme simply identifies a discontinuity, 
the gas-kinetic scheme tries to resolve it in a physically meaningful way.  

Fig. \ref{fig:delerystressuv} shows the distribution of turbulent stress $\tau_{xy}$. The agreement with experimental values (e.g. the PIV investigation by Sartor \cite{sartor2012piv}, not shown here)  is only qualitatively acceptable (the maxima are about $20\%$ apart). 

The degree of rarefaction  is shown in Fig. \ref{fig:deleryrarefactionlevel}: high values, in the same order as the ones observed in the airfoil flow cases, appear in the proximity of the shocks and might be at the origin of the different behavior of the two schemes. 


\subsubsection{Flow Case 2. Supersonic compression corner at high Reynolds number}
\label{subsubsec:settles}

This flow case concerns the well known experiments carried out by Settles \cite{settles1976details}.
The experiments include four ramp angles: $8^\circ$, $16^\circ$, $20^\circ$ and $24^\circ$. The smallest angle is not sufficient to separate the boundary layer, the $16^\circ$ ramp is just enough to cause an incipient separation, whereas separation occurs with the two steeper angles.
Whereas conventional schemes seem able to simulate the first two cases (also the computations published by Settles in 1979 \cite{settles1979detailed} were in good agreement with experiments), the separation area is systematically underpredicted. 

\begin{figure}[h!]
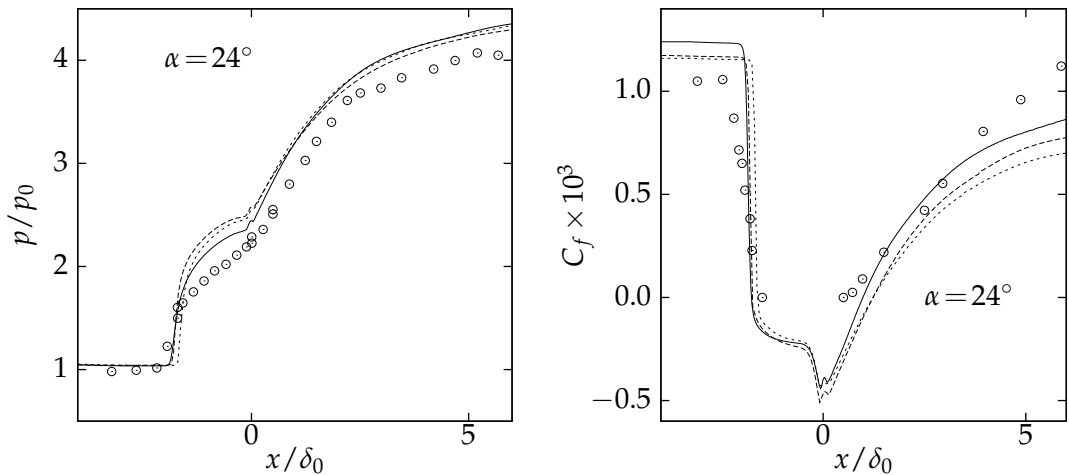

  \begin{center}
    \input{cpramp24gridconv_}
    \input{cframp24gridconv_}
  \end{center}
  \caption{Flow Case 2. Grid convergence. 
   (freestream conditions: $M = 2.85$, $Re = 7.0\times 10^7$ per length unit, $\delta_0 = 0.023\,m$).
   (\solidrule)  Gas-kinetic scheme (GKS) on finest grid,
   (\protect\thindashedrule) GKS on medium grid (only shown for $\alpha = 24^\circ$), 
   (\protect\verythindashedrule) GKS on coarsest grid (only shown for $\alpha = 24^\circ$),  
   (\protect\fewdots): experimental data from Settles \cite{settles1979detailed}.    
  }
  \label{fig:cpramp24settlesgridconv}
\end{figure}

\begin{figure}[h!]
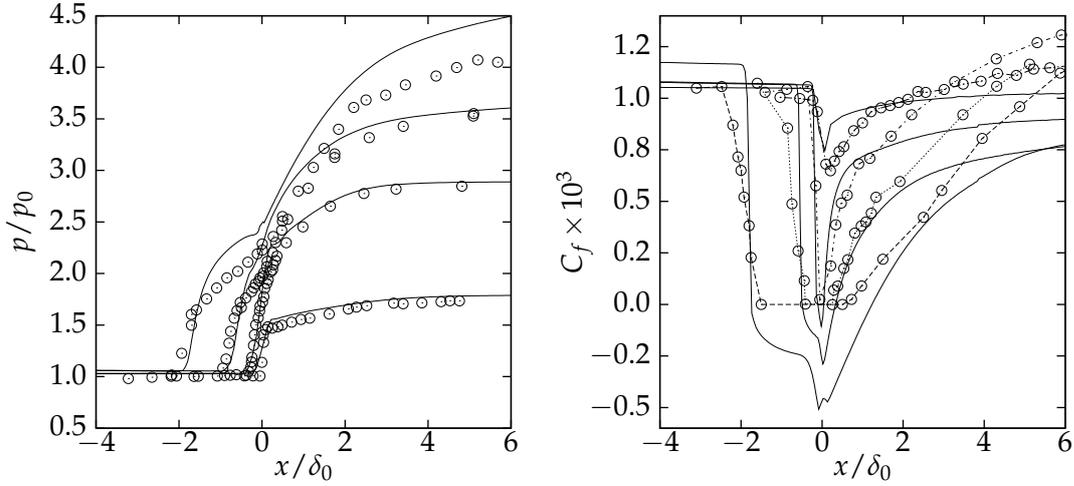

  \begin{center}
    \input{cprampall_}
    \input{cframpall_}
  \end{center}
  \caption{ Flow Case 2. Pressure and friction coefficient calculated for four different compression corner flows, with angles of $8^\circ$, $16^\circ$, $20^\circ$ and $24^\circ$ (freestream conditions: $M = 2.85$, $Re = 7.0\times 10^7$ per length unit, $\delta_0 = 0.023\,m$).
   (\solidrule) GKS on finest grid,
   (\protect\fewdots): experimental data from Settles \cite{settles1979detailed}.  
  }
  \label{fig:cpallrampssettles}
\end{figure}

Pressure and skin friction coefficient obtaind with the turbulent gas-kinetic scheme for Settles' experiment are shown in Fig. \ref{fig:cpallrampssettles}. The evidence of grid convergence is shown in Fig. \ref{fig:cpramp24settlesgridconv} for the $24^\circ$ ramp. For all ramp angles, the good agreement in the interaction area is evident whereas the flow after-reattachment is  predicted with lower accuracy. 
A likely cause is related to the inadequacy of the underlying $k$-$\omega$ model to separated flow in supersonic regime. 

A further aspect investigated in this flow case is the sensitivity to Reynolds number. 
In particular, the upstream interaction length has been calculated at different Reynolds numbers (for the $24^\circ$ ramp), the results are summarized in Table \ref{tab:settlesreynolds}. 
Interestingly, the lengths calculated seem to fit acceptably into the empirical relation devised by Settles (\cite{delery1986shock}) for flows at $M\simeq3$:

\begin{equation}
\left( \frac{L_0}{\delta_0} \right) \, Re_{\delta_0}^{1/3} = 0.9 e^{0.23 \alpha},
\label{eq:settlesempirical}
\end{equation} 

\noindent where  $L_0$ is the upstream influence length and $\delta_0$ is the boundary layer thickness. 

\begin{table}[h!]
\begin{center}
\begin{tabular}{rrrr}
\hline \hline
$Re_{\delta_0}$     & {$L_0/\delta_0$}   & {$\left( \frac{L_0}{\delta_0} \right) \, Re_{\delta_0}^{1/3}$}   &  Error ($\%$) \\ 
\hline
$    740,000$ & 2.38           &   216.74   &  3.53   \\ 
$1,694,000$ & 1.98           &   236.34   &  5.19     \\ 
$2,120,000$ & 1.70           &   218.15   &  2.90    \\ 
$2,912,000$ & 1.63           &   233.42   &  3.90     \\ 
\hline\hline
\end{tabular}
\end{center}
\caption{Flow Case 2. Compression corner, $M=2.85$, $\alpha = 24^\circ$. Sensitivity to Reynolds number. Empirical law  $0.9 e^{0.23 \alpha}$, refer to \cite{delery1986shock}.}  
\label{tab:settlesreynolds}
\end{table}

\begin{figure}[h!]
\begin{center}
\includegraphics[width=128mm]{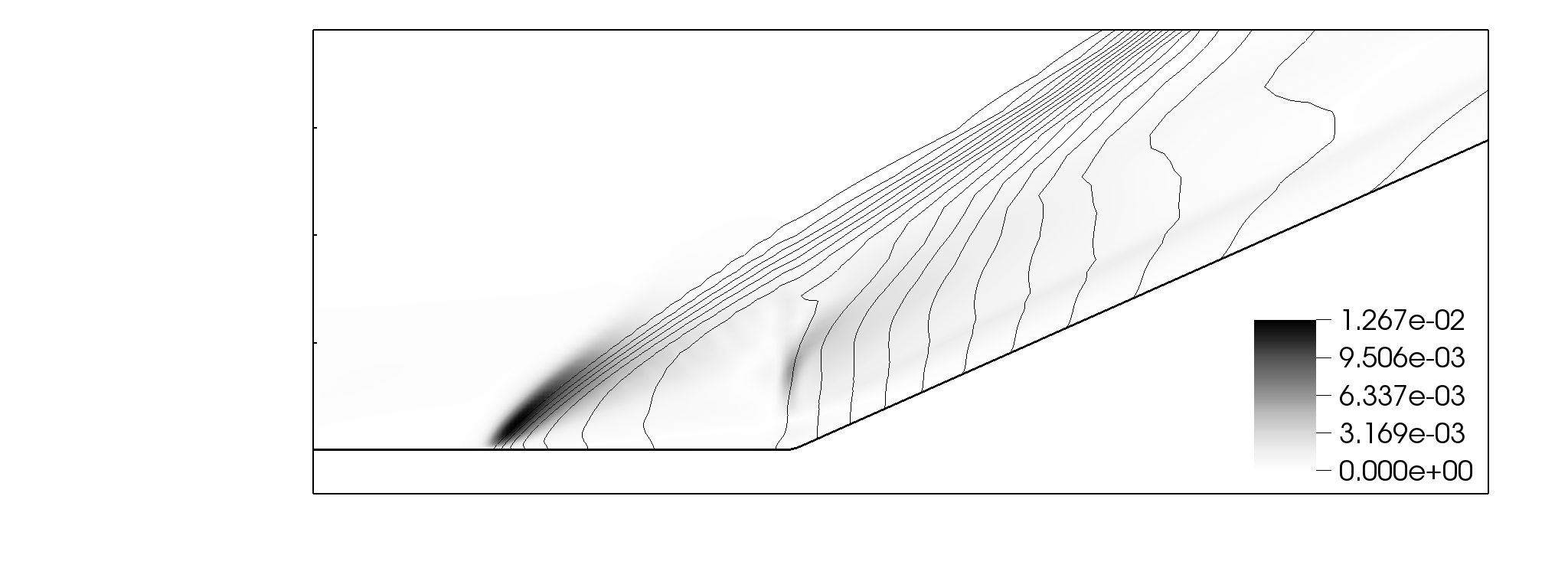}
\end{center}
\caption{Flow Case 2 $\alpha=24^\circ$. Degree of rarefaction according to Eq. \ref{eq:epsilonkomega} in supersonic compression corner flow, $M=2.85$ (iso-contours in grayscale). 20 pressure contour lines have been added for reference. }
\end{figure}

\subsubsection{Flow Cases 3 and 4. Supersonic compression corner and impinging shock at low Reynolds number}
\label{subsubsec:bookey}

Both flow cases refer to the experiments carried out by Bookey \cite{bookey2005experimental}, devised especially to provide a benchmark case for LES and DNS. The first experiment concerns a ramp with an angle of $24^\circ$, the second an impinging shock generated by a ramp with an angle of $12^\circ$.

\begin{figure}[h]
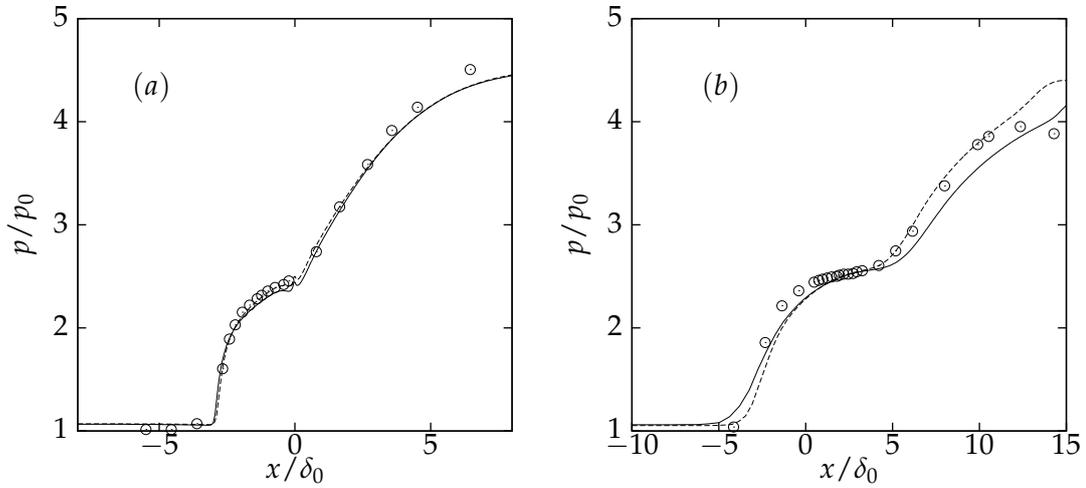

  \begin{center}
    \input{cp24bookey_}
    \input{cpRSbookey_}
  \end{center}
  \caption{
  Flow Cases 3 and 4. Distribution of static pressure calculated for the compression corner (a) with an angle of $24^\circ$ and for the reflected shock (b) originating form a compression corner at $12^\circ$.  Freestream conditions (both flows) are $M=2.90$, $Re_{\theta} = 2\; 400$.
   (\solidrule) GKS on grid 1,
   (\protect\verythindashedrule) GKS on grid 2,  
   (\protect\fewdots): experimental data from Bookey \cite{bookey2005experimental}.
   Grid sizes compression corner: $384 \times 192$, $512 \times 168$ respectively, grid size impinging shock: $384 \times 208$, $496 \times 304$ respectively.  Grid 1 and 2 have different resolution and have been generated with different algorithms.
  }
\label{fig:cpbookey}
\end{figure}

\begin{figure}[h!]
\begin{center}
\includegraphics[width=128mm]{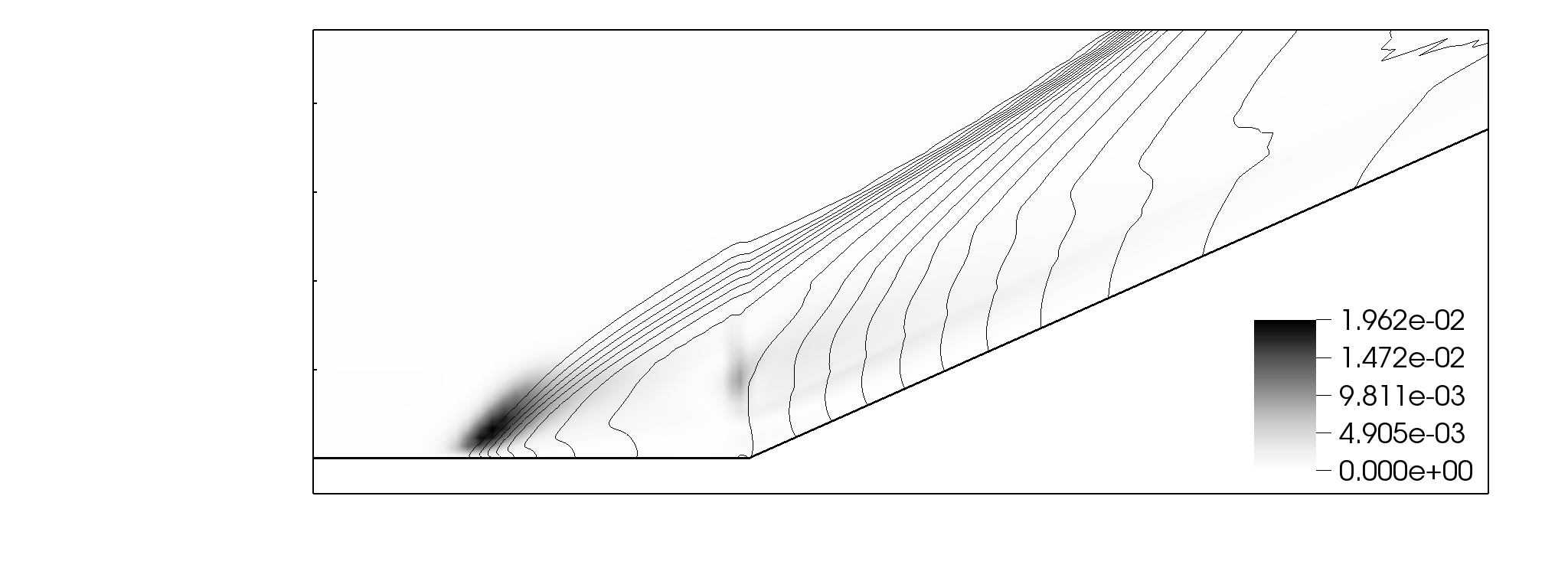}
\end{center}
\caption{Degree of rarefaction according to Eq. \ref{eq:epsilonkomega} in  Flow Case 3, supersonic compression corner, $M=2.90$ (iso-contours in grayscale). 20 pressure contour lines added for reference. }
\label{fig:bookeyknt}
\end{figure}

Fig. \ref{fig:cpbookey} shows the pressure distributions for both the compression corner and the impinging shock, which are in reasonably good agreement with the experiments. Fig. \ref{fig:bookeyknt} shows the distribution of the degree of rarefaction in the case of the compression corner, which reaches significant values across the shock wave and on the slip line. 

Table \ref{tab:reynoldstrendbookey} summarizes the evaluation of the upstream influence distance - in the case of the compression corner -  as a function of Reynolds  number, comprised in a range between 1900 and 5000. 
The trend is compared with Eq. \ref{eq:settlesempirical}, the scaling law proposed by Settles \cite{delery1986shock} used in Flow Case 2. In particular, the relation between the upstream interaction length and the (displacement thickness based) Reynolds number can be empirically related to Eq. \ref{eq:settlesempirical} with the exponent changed from $0.23$ to $0.1885$:  $\left( {L_0}/{\delta_0} \right) \, Re_{\delta_0}^{1/3} = 0.9 \exp{\left(0.1885 \alpha\right)}$.

\begin{table}[h!]
\begin{center}
\begin{tabular}{rrrr}
\hline \hline
$Re_{\theta}$     & {$L_0/\delta_0$}   & {$\left( \frac{L_0}{\delta_0} \right) \, Re_{\delta_0}^{1/3}$}   &  Error ($\%$) \\ 
\hline
$1\,900$ & 2.70           &   80.72   &  2.03   \\ 
$2\,300$ & 2.58           &   81.85   &  0.65     \\ 
$3\,200$ & 2.35           &   83.59   &  1.45    \\ 
$5\,900$ & 1.92           &   84.08   &  2.05     \\ 
\hline\hline
\end{tabular}
\end{center}
\caption{Flow Case 3. Compression corner, $M=2.90$, $\alpha = 24^\circ$. Upstream influence distance  as a function of Reynolds number. Empirical law adjusted to $0.9 e^{0.1882 \alpha}$ (original: $0.9 e^{0.23 \alpha}$, refer to \cite{delery1986shock}).}  
\label{tab:reynoldstrendbookey}
\end{table}

\subsubsection{Flow Case 5. Supersonic impinging shock at low Reynolds number }
\label{subsubsec:dupont}

\begin{figure}[h!]
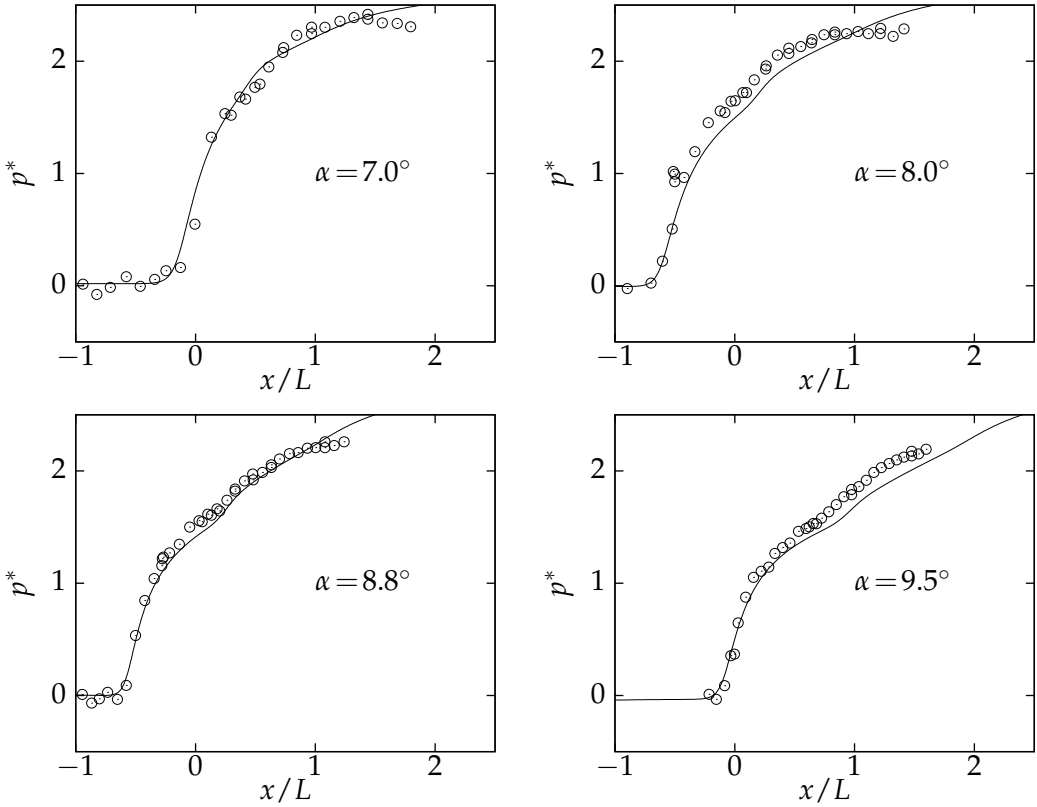

  \begin{center}
    \input{cpRS70dupont_}
    \input{cpRS80dupont_}
    \input{cpRS88dupont_}
    \input{cpRS95dupont_}
  \end{center}
  \caption{Flow Case 5. Reflected shock wave at low Reynolds number and Mach $M=2.3$ (experimental data from \cite{dupont2006space}) with angles ranging from $7.0$ to $9.5$ degrees. $p^* = (p - p_1)/(p_2 - p_1)$, where $p_1$ and $p_2$ are the  inviscid pressure values upstream and downstream of shock, respectively. The interaction length $L$ is measured from the foot of the reflected shock to the reattachment point. Results obtained on a grid with size: $496 \times 304$. }
\label{fig:cpcfdupont}
\end{figure}

\begin{figure}[h!]
\begin{center}
\includegraphics[width=128mm]{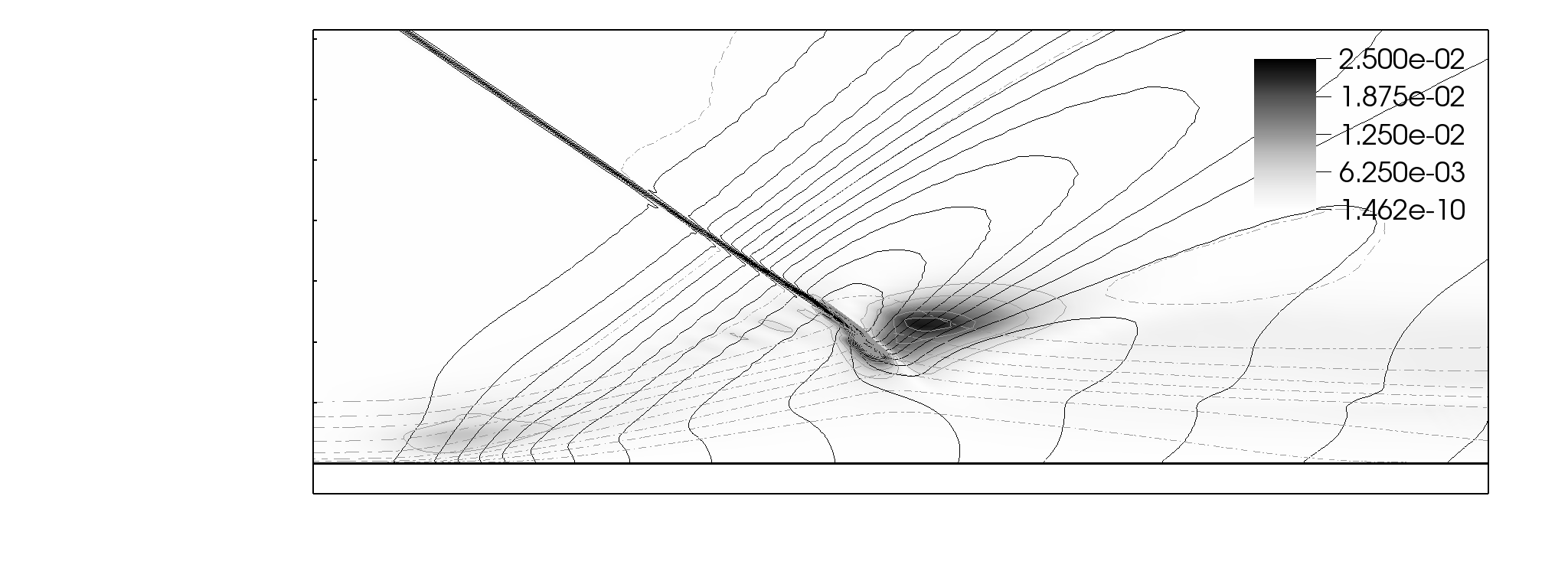}
\end{center}
\caption{Degree of rarefaction according to Eq. \ref{eq:epsilonkomega} in Flow Case 5, reflected shock, $M=2.30$ (iso-contours in grayscale). 20 pressure contour lines have been added for reference.  }
\label{fig:dupontknt}
\end{figure}

The experiment by Dupont \cite{dupont2006space} is similar to flow case 3, carried at a Mach of $2.3$ and a slightly different Reynolds number. 
It has the advantage of providing results for four different angles: $7^\circ$, $8^\circ$, $8.8^\circ$ and $9.5^\circ$ respectively. 
The pressure distribution is reported in Fig. \ref{fig:cpcfdupont} and shows a reasonably good agreement with the experiment. 
The degree of rarefaction, shown in Fig. \ref{fig:dupontknt}, reaches significant values at the foot of the impinging shock and at the separation point.

\subsubsection{Flow Case 6. Supersonic compression corner at Mach $5$ }
\label{subsubsec:hypersonic}

\begin{figure}[h!]
  \begin{center}
    \input{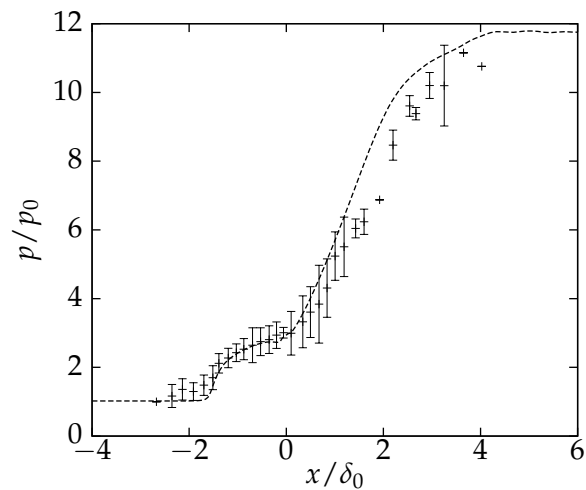}
  \end{center}
  \caption{Compression corner $ M=5$, Flow Case 6. Experimental values from Dolling {\sl et al.} \cite{dolling1991unsteady}. Results obtained on a grid with size: $384 \times 160$}
\label{fig:cpcfm5}
\end{figure}

\begin{figure}[h!]
\begin{center}
\includegraphics[width=128mm]{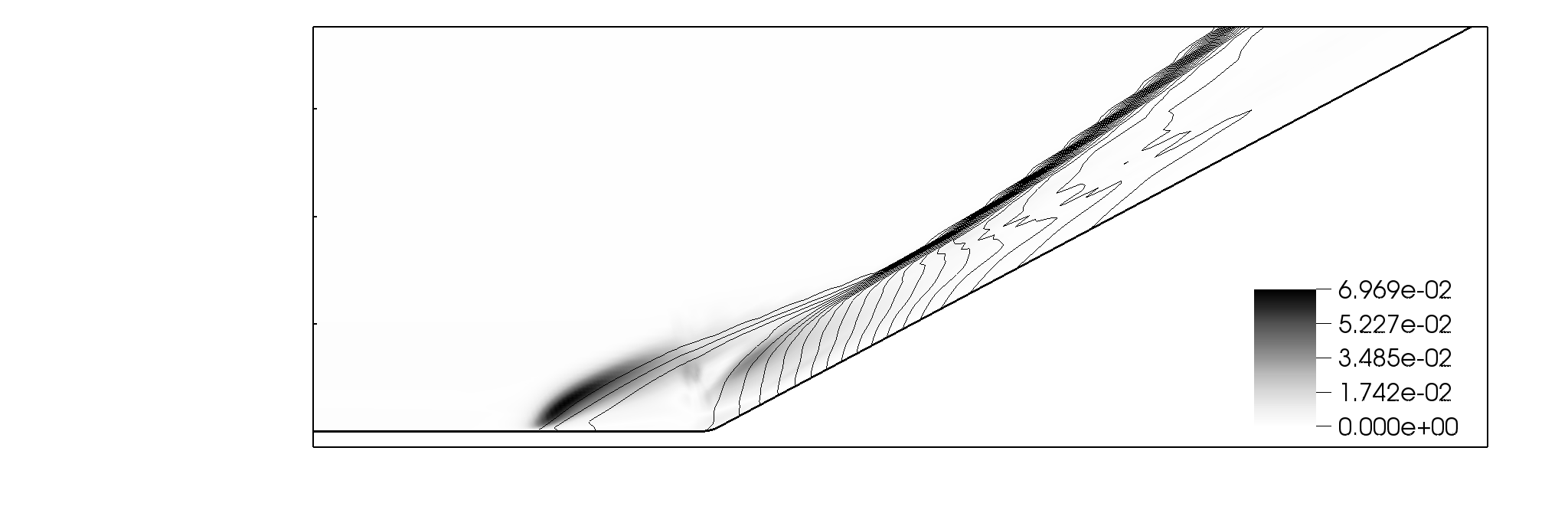}
\end{center}
\caption{Degree of rarefaction according to Eq. \ref{eq:epsilonkomega} in Flow Case 6, supersonic compression corner, $M=4.95$ (iso-contours in grayscale).  20 pressure contour lines have been added for reference. }
\label{fig:dollingknt}
\end{figure}

This flow case, investigated by Dolling {\sl et al.} \cite{dolling1991unsteady}, has been selected to extend the mach Number range to the borders of hypersonics. This flow case has been however calculated assuming adiabatic wall conditions. 
Results from RANS and hybrid simulations can be found in Edwards {\sl et al.} \cite{edwards2008numerical}. 
In Fig. \ref{fig:cpcfm5} the pressure distribution predicted by the turbulent gas-kinetic scheme is compared to the experimental values by Dolling {\sl et al.} \cite{dolling1991unsteady}, highlighting an acceptable agreement. 
Fig. \ref{fig:dollingknt} shows the distribution of the degree of rarefaction, which reaches values as high as $0.07$. 

\section{Conclusions}
\label{sec:conclusions}

The predictions obtained with the turbulent gas-kinetic scheme are surprisingly close to the experimental value, at least in terms of mean values and shock position.
The turbulent gas-kinetic scheme does not correct in any way the evaluation of the two turbulent quantities, $k$ and $\omega$, provided by the standard model.  However, the interaction with the shock is handled, in all examples,  in a much more accurate and convincing way. The degree of rarefaction, measured in terms of timescales ratio,
reaches in shocklayers values  well above the boundary of the continuous regime and definitely in the transitional regime. 
It is precisely in these ``rarefied'' flow regions that the gas-kinetic solution differs from the solution obtained with conventional schemes and shows a better agreement with experimental data. Incidentally, rarefaction increases with Mach and decreases with Reynolds. Potentially, the gas-kinetic scheme seems to have the potential to improve predictions in special flow case such as hypersonic flight.

Whereas the turbulent gas-kinetic scheme 
provides merely a higher accuracy in smooth flow, in the flow regions where the degree of rarefaction reaches significant values, it seems able to 
deviate from conventional schemes and in so doing
provide physically more consistent solutions. 
The analysis presented in this paper reveals that the turbulent gas-kinetic scheme includes a rarefaction ``sensor'' which activates the deviations from the conventional schemes. These  corrections terms are modeled by the underlying gas kinetic theory and, unlike  conventional turbulence modeling, 
do not require any assumption on the nature of the turbulence nor 
any series expansion  of the turbulent stress tensor. 

The properties of turbulent gas-kinetic schemes are still largely unknown. It might be useful to 
investigate the influence of the main parameters such as order of the Chapman-Enskog expansion for $f_0$, reconstruction order, time integration and pre-conditioning.
Further, validation is still at its beginnings: three-dimensional flow cases are still unexplored and so is the role played by the choice of the platform model. 

An extension to unsteady simulations and in particular to Large Eddy Simulation or to RANS-LES hybrid techniques would be possible without any significant change to the scheme, except for the evaluation of the subgrid turbulent relaxation time. The turbulent gas-kinetic scheme might  even be more suitable to LES, since,  as pointed out by Chen {\sl et al.} \cite{chen2004expanded}, the largest unresolved scale of motion is normally not much smaller than the smallest resolved scale, leading to large $\tau/\widehat{\tau}$ or $\tau/\Delta t$ ratios.




\bibliographystyle{plain}   
\bibliography{/home/rigm/zhaw/Papers/AIP/bibliobgk}


\end{document}